\documentclass[aps,prl,groupedaddress,amsmath,amssymb,reprint,superscriptaddress]{revtex4-1}
\usepackage{amsmath,amssymb}
\usepackage{graphicx}
\usepackage{physics}
\usepackage[T2A]{fontenc}
\usepackage[cp1251]{inputenc}
\begin{document}
%
%\title[Short title: Comment on half-integer quantum numbers for the total angular momentum of
%photons in light beams with finite lateral extensions]%
\title{Comment on half-integer quantum numbers for the total angular momentum of
photons in light beams with finite lateral extensions%
}
%
%\author[Short authors list]{M. F{\"a}hnle$^{*}$ \refaddr{label1}}
%       M. Haag\refaddr{label1}}
%\addresses{
%\addr{label1}Max Plank Institute for Intelligent Systems, Heisenbergstr. 3, 70569 Stuttgart,\\  $*$corresponding author, E mail: faehnle@is.mpg.de
%}

\author{M. F{\"a}hnle}
\email[Corresponding author: ]{faehnle@is.mpg.de}
\affiliation{Max Plank Institute for Intelligent Systems, Heisenbergstr. 3, 70569 Stuttgart, Germany}

%
%% or for single author or if all authors are from the same institute:
%
%  \author[Short authors list]{1st Author, 2st Author, \ldots}
%  \address{Institute}
%
%% Fields in square brakets (short title and short authors list) are
%% optional. Use them if your entries exceeds 45 characters.
%

%\sloppy

\begin{abstract}
Recently the spectacular result was derived quantum mechanically that the
total angular momentum of photons in light beams with finite lateral
extensions can have half-integer quantum numbers. In a circularly
polarized Gauss light beam it is half of the spin angular momentum which
it would have in a respective infinitely extended wave. In another paper
it was shown by a classical calculation that the magnetic moment induced
by such a beam in a metal is a factor of two smaller than the one induced
by a respective infinitely extended wave. Since the system's angular
momentum is proportional to its magnetic moment it could be assumed that
the classical result for the magnetic moment reflects the transfer of the
total angular momenta of the beam photons to the metal. Here we show that
there is no hint that this is indeed the case.

%
%\keywords Up to six keywords
%

\end{abstract}
\pacs {}
\maketitle

\section{Introduction}
The angular momentum of photons is heavily discussed in many fields of
optics. Starting point thereby often is \cite{1f} that Maxwell's equations are
 invariant under rotations around any direction. 
This yields the  conserved quantity $L_{i} + S_{i}$, where $L_{i}$ and $S_{i}$ are the i-components of the
orbital and the spin angular momentum, and where the i-axis is parallel to the
 respective axis of rotation. This quantity is usually conceived as the
 total angular momentum of photons. The quantum numbers for this total
 angular momentum are integers.
    K.E. Ballantine et al.  \cite{2f} considered the realistic situation in
 which there is a beam of light with finite lateral extension. Then the system
 is no longer invariant against rotations around any direction, in the therein mentioned special case of 
a paraxial beam the reduced symmetry is given by
the rotations around the axis of the beam which we denote as z-axis.
 In such a paraxial beam both spin and orbital angular momentum are valid and 
 independent, which can even be proven for more general cases \cite{xtra}. %quelle
Therefore one can discuss a general linear combination,
\begin{align}
\widehat{J}_{z,\gamma} =\widehat{L}_{z} + \gamma \widehat{S}_{z},
\end{align}
where $\widehat{\mathbf{L}}$ and $\widehat{\mathbf{S}}$ are the operators
  of the orbital and spin angular momentum, respectively.
The operator $\widehat{\mathbf{J}}_{z,\gamma}$
 generates an observable. 
If the electric field of the beam is an
eigenfunction, $\{\widehat{L}_{z} + \gamma \widehat{S}_{z}\} E = j_{\gamma} E$ the observable can
be considered having the meaning of a total angular momentum. 
In \cite{2f} it was shown that $\gamma$ and $j_{\gamma}$ are either both integers or both half-integers.
The first type includes the usual angular momentum operator $\widehat{L}_{z} + \widehat{S}_{z}$,
 with eigenvalues which are integer quantum numbers. 
The second type,
 typified by $\{\widehat{L}_{z} +\frac{1}{2}\widehat{S}_{z}\}$, corresponds to half-integer quantum
 numbers. 
The electric fields of circularly polarized Gauss-Laguerre
 beams  \cite{3} are  eigenfunctions of $\{\widehat{L}_{z} + \gamma \widehat{S}_{z}\}$,  
and their photons have total angular momenta with half-integer quantum numbers \cite{2f}. 
For a circularly polarized Gauss beam (which is non-helical) the photons have
 a zero orbital angular momentum and a spin angular momentum $S_{z}$ which is
 the  spin angular momentum $s_{z}$ which they would have in a respective
infinitely extended wave, i.e., $S_{z} =  s_{z}$. 
As shown in Ref.~\cite{2f} the total angular
 momentum of photons in a circularly polarized Gauss beam is $\frac{s_{z}}{2}$. 
In Ref.~\cite{2f}  the half-integer quantization of the total angular momentum of beam
  photons was demonstrated by noise measurements.

 In  Ref.~\cite{4} it was shown by a classical treatment of the inverse Faraday
 effect (using Newton's equation of motion for the movement of electrons
 in a classical electric field described by Maxwell's equations) that the
 magnetic moment $\mathbf{M}$ induced in a metal by a circularly
polarized Gauss beam is a factor of two smaller than the one, $\mathbf{M}_{\text{g}}$,
 induced by an infinitely extended circularly polarized wave. 
The reduction is caused by a drift current occurring as a result of the
 finite lateral extension of the Gauss beam, which contributes by 
$\mathbf{M}_{\text{d}} = \frac{-\mathbf{M}_{\text{g}}}{2}$ to the total magnetic moment 
$\mathbf{M} = \mathbf{M}_{\text{g}}+\mathbf{M}_{\text{d}} =\frac{\mathbf{M}_{\text{g}}}{2}$. 
The effect of an electromagnetic wave on matter reflects the properties of the wave in detail,
being more or less a 'fingerprint' of the wave. %präziser
The system's angular
 momentum is proportional to its magnetic moment, and the above discussed
reduction of the magnetic moment reflects a reduction of the system's
angular momentum by a factor of two. 
The angular momentum of the system
 is physically generated by a transfer of angular momentum from the beam to the
 metal.
Therefore it could be assumed that the reduction of
 the system's angular momentum by a factor of two is related to the
 half-integer quantum number of the total angular momentum of the photons
 of the beam. 
We now show that there is no hint that this is actually the
 case. This is very important for the community working on the total
 angular momentum of light beams and of their photons, and for the
 community working on the inverse Faraday effect and on all-optical
 switching of the magnetization of a material \cite{5}.

\section{ORBITAL and SPIN ANGULAR MOMENTUM of a GAUSS BEAM}

\subsection{ Calculation for classical fields}

 The central idea is that in the classical treatment of  Ref.~\cite{4} the Gauss
 beam is not described by a quantum electrodynamics in which the
 electromagnetic  field is built by photons, but it is described by a classical electric
 field $\mathbf{E}\left(\mathbf{r},t\right)$. Therefore the electrons of the metal do not 'know' the
 half-quantization of the photons of the beam, but they 'see' a classical
 Gauss beam. Therefore we now calculate the spin and orbital angular
 momentum of this classical field. The angular momentum of an
 electromagnetic field is \cite{6f}
\begin{align}
\mathbf{J} = \epsilon_{0} \int d^{3}r \mathbf {r}\times \left(\mathbf{E}\times\mathbf{B}\right).    
\end{align}
 This may be subdivided \cite{9} in a spin angular momentum $\mathbf{S}$ and an orbital
 angular momentum $\mathbf{L}$, $\mathbf{J}$ = $\mathbf{L}$+$\mathbf{S}$. 
We now calculate $\mathbf{S}$ and $\mathbf{L}$ for a classical
 circularly polarized Gauss beam. Afterward we consider the total angular
 momentum of photons in such a beam and explain it in a simple view based
 on the 'trajectories' of the photons.

     We consider the electric field $\mathbf{E}$ of a circularly polarized Gauss
 beam
 with width $w$ and angular frequency $\omega$,

 \begin{align}                                                                    
 \mathbf{E}= \frac{E_{0}}{\sqrt{\left(2 \pi^{3/2}\right) w^{3}}} \exp{-r^2/w^2}\exp{-i \omega t}\begin{pmatrix} 1 \\ \pm i \\0 \end{pmatrix},
\end{align}

 where (+) and (-) stand for left and right circular polarization. 
This ansatz is normalized in the sense that the integral over the space of
 $\mathbf{E}^{\dagger} \mathbf{E}$  yields  $E_{0}^2$. 
In  Ref.~\cite{4} an ansatz without the prefactor $\frac{1}{\sqrt{\left(2
 \pi\right)^{3/2} w^3}}$ is used. We take the normalized form because we compare
 our results with those obtained for an infinitely extended circularly  polarized wave,
\begin{align}                                   
 \mathbf{E} = \frac{E_{0}}{\sqrt{V}} \exp{-i \omega t}\begin{pmatrix} 1 \\ \pm i \\0 \end{pmatrix}   ,
 \end{align}                                   
 which is also normalized in the above sense. 
Of course the basic result  of \cite{4}, 
$\mathbf{M} = \frac{\mathbf{M}_{g}}{2}$, obtained for the non-normalized ansatz in \cite{2f} also holds 
 when using the normalized ansatz.

 For the classical spin angular momentum \cite{Barnett2010},

\begin{align}
 \mathbf{S} =\frac{\epsilon_{0}}{\left(2 i \omega\right)} \int d^{3}r  \mathbf {E}^{\dagger}\times\mathbf{E}   
\end{align}
 we get both for the Gauss beam, $S_{z}$, and for the infinitely extended
 wave, $s_{z}$,  $\mathbf{S}=\left(0,0,S_{z}\right)$ with
\begin{align}
 S_{z} = \pm \epsilon_{0} E_{0}^{2}/\omega.    
\end{align}
 The same result, $S_{z}$ = $s_{z}$, was also found for the photons in the Gauss  beam.

 The orbital angular momentum \cite{Barnett2010}
\begin{align}
 \mathbf{L} = \frac{\epsilon_{0}}{(2 i \omega)} \sum_{i=x,y,z} \int d^{3} r
       (E_{i})^{*} \left(\mathbf{r}\times\grad\right) E_{i}     
\end{align}
 we get for the infinitely extended wave and for the Gauss beam the
 result $ \mathbf{L}=0$. 
Remember that the orbital angular momentum of photons in the Gauss
 beam is also zero, and it is of course also zero for an infinitely
 extended wave.

 Altogether this means that the orbital angular momentum  of the  classical
 circularly polarized Gauss beam is zero, as the corresponding quantity
 of  the photons in that beam, and that the spin angular momentum $S_{z}$ of the
 beam is equal to the spin angular momentum $s_{z}$ of the infinitely
 extended  wave. 

However, the total angular momentum of the
 photons in the beam is $\frac{S_{z}}{2}$. Therefore there is no hint that the result
 $\mathbf{M}=\frac{\mathbf{M}_{g}}{2}$ of the classical calculation can be related to the transfer of
 the  total angular momentum of the photons in the beam.

 The question therefore arises whether the calculation of the magnetic
 moment induced by the Gauss beam should be done quantum mechanically
 instead classically. 
However, the length- and energy-scales occurring in
 that problem are withing the validity regime of the classical theories
 (Newton's mechanics and Maxwell's theory), and therefore a quantum
mechanical treatment (which would be extremely complicated) is not
 necessary. 
There is no doubt that the results of \cite{4} are correct. All
 this demonstrates that the factors $\frac{1}{2}$ appearing in the theories of  Ref.~\cite{2f} and
 Ref.~\cite{4}  do not have the same physical basis.

\subsection{Comments on the orbital angular momentum of photons}

In a Gauss beam the z-component $L_{z}$ of the orbital angular momentum is a
 conserved quantity. We now want to explain the result of  Ref.~\cite{2f} that the
 orbital angular momentum $L_{z}$ of photons in the non-helical Gauss beam is
 zero. We explain it from a simple viewpoint which is based on the notion
 of 'trajectories' of the photons. In principle, the photon state is
 described by a wavefunction so that one cannot associate a well-defined
 position to a photon. However, we can construct a wavepacket by a
 superposition of waves with different wavevectors, and this wavepacket
 can  be made well localized in space. 
The lateral extension of a helical beam
 is much larger than the wavelength of the light, and therefore problems
 arising from the position-momentum uncertainty are not important, and
 the  concept of a photon trajectory seems  to be not too bad.
 For a particle with a trajectory, i.e., for which we can simultaneously
 define a position $\mathbf{r}$ and a linear momentum $\mathbf{p}$, the orbital angular
 momentum
 is $\mathbf{l} = \mathbf{r}\times\mathbf{p}$. The trajectories of the photons in a Gauss beam are parallel
 to the beam axis, i.e.,$ \mathbf{p} = \left(0,0,p\right)$. This leads directly to $l_{z}$ = 0.

To consider the photon trajectory in a helical beam we note that the
 trajectory may be conceived as the saddle-point solution of a
 path-integral representation of the wavemechanics, and this is given by
 the Eikonal equation. 
This equation says that the photons choose the
 path  between two points in space for which they have the shortest running
 time.
 In a homogeneous medium this is always the straight path (principle of
 Fermat). For a helical beam we do not have homogeneity. 
So the  'shortest'  path is not a straight line. 
When we consider the 'shortest' path as the
 trajectory of the photon, then the orbital angular momentum $l_{z}$ of the
 photon is  $ \hbar m$,where m describes the phase winding of the helical
 wavefront according to the complex exponential $\exp{i m/\theta}$, i.e., we
 now have a non-vanishing orbital angular momentum of the photon, in
 contrast to the zero orbital angular momentum of the photon in a
 non-helical beam.

\section{Conclusion}

 A recent quantum mechanical calculation \cite{2f} has given the spectacular
 result that the total angular momentum of photons in circularly
 polarized
 light beams can have half-integer quantum numbers. Another recent
classical treatment \cite{4} has given an also spectacular result, namely
 that
 the magnetic moment induced by such a beam in a metal is a factor of two
 smaller than the one induced by a respective infinitely extended wave.
 We
demonstrated in the present paper that the two results, which seem to
 have
 the same underlying physics, have nothing to do with each other in
 reality. The result of the quantum mechanical treatment that the photons
 in a non-helical Gauss beam have zero orbital angular momentum whereas
 they have a non-zero orbital angular momentum in a helical light beam
 could be explained from a simple viewpoint based on the notion of photon
 trajectories. Thereby the meaning of a photon trajectory in a light beam
 with a width which is much larger than the wavelength of the light is
 discussed.
%\\
%\strut\hfill
%\verb|\section*{Acknowledgments}|%
%\hfill\strut%

%\\
\section*{Acknowledgements}
The author is indebted to Riccardo Hertel and Paul
 Eastham for many interesting discussions.
\\

\bibliography{extra}

\end{document}